\documentclass[prl,twocolumn,showpacs,amsmath,amssymb,superscriptaddress]{revtex4-1}
\usepackage{graphicx}
\usepackage{dcolumn}
\usepackage{bm}
\usepackage{natbib}
\usepackage{amssymb}
\usepackage{dcolumn}
\usepackage[usenames,dvipsnames]{xcolor}

\newcommand{\nc}{\newcommand}

\nc{\be}{\begin{equation}} \nc{\ee}{\end{equation}}
\nc{\bea}{\begin{eqnarray}} \nc{\eea}{\end{eqnarray}}
\nc{\bean}{\begin{eqnarray*}} \nc{\eean}{\end{eqnarray*}}

\newcommand{\comments}[1]{}

\begin{document}

\title{Full first-principles theory of spin relaxation in group-IV materials}
\author{O. D. Restrepo}
\affiliation{Dept.\ of Materials Science and Engineering, The Ohio State University, Columbus, OH 43210}
\author{W. Windl}
\affiliation{Dept.\ of Materials Science and Engineering, The Ohio State University, Columbus, OH 43210}

\date{\today}
\begin{abstract}
We present a generally applicable parameter-free first-principles method to determine electronic spin relaxation times and apply it to
the technologically important group-IV materials silicon, diamond and graphite. We concentrate on the Elliott-Yafet mechanism, where spin relaxation is induced by momentum scattering off phonons and impurities.
 In silicon, we find a $\sim T^{-3}$ temperature dependence of the phonon-limited spin relaxation time T$_1$ and a value of 4.3 ns at room temperature, in agreement with experiments. For the phonon-dominated regime in diamond and graphite, we predict a stronger $\sim T^{-5}$ 
 and $\sim T^{-4.5}$ dependence that limits $T_1$ (300 K) to 180 and $5.8$ ns, respectively.
A key aspect of this study is that the parameter-free nature of our approach provides a method to study the effect of {\em any} type of impurity or defect on spin-transport. 
Furthermore we find that the spin-mix amplitude in silicon does not follow the $E_g^{-2}$ band gap dependence usually assigned to III-V semiconductors but follows a much weaker and opposite $E_g^{0.67}$ dependence. This dependence should be taken into account when constructing silicon spin transport models.   
\end{abstract}
\maketitle

The physical roadblocks looming in the charge-based semiconductor device technology require paradigm-shifting approaches to create new logic devices capable of lower power consumption and higher performance. This has motivated a search for new alternative logic variables, among which the spin of electrons is a natural candidate, which needs to be efficiently and reliably injected, transported, and detected. Although extensive studies have been done in direct-gap materials, the understanding of spin-transport and spin-life time dependence in the technologically relevant group-IV materials is surprisingly incomplete. Silicon and the carbon polytypes diamond and graphite are particularly relevant because long spin relaxation times can be expected in materials with inversion symmetry and low atomic number $Z$. For those, the main spin-relaxation mechanism at high temperatures is the Elliott-Yafet (EY) mechanism mediated through spin-orbit coupling \cite{Elliott-1954,Yafet-1963}, which scales as ~Z$^2$. 
  
Silicon, an attractive potential spintronics material due to its compatibility with current technologies,  
has a relatively large spin orbit-coupling (44 meV) \cite{Ferry-2001}. Nevertheless,  Lepine's electron spin resonance (ESR) measurements \cite{Fabian-2007,Lepine-1970}, recently confirmed for low temperatures by Appelbaum et al.
\cite{Appelbaum-H-M-2007,Huang-M-A}, found for its spin relaxation time $T_1$ a value of 7 ns at room temperature, which puts it well within the usable range. The experimental situation is less clear for carbon, whose lower $Z$-number promises longer spin relaxation times. 
Diamond with a spin-orbit coupling of 13 meV \cite{Carrier-W-2004} is especially expected to have a long electronic spin relaxation time, which however has never been measured.  For graphite, the most recent experimental data from 1961 \cite{Wagoner-1960,Muller-1961} suggest a large range for $T_1$  between 1-300 ns.
For both diamond and graphite, no reliable theoretical predictions have been reported. Finally, the $E_g^{-2}$ dependence of the spin-mix amplitude found for III-V semiconductors \cite{Z-F-DS-2004} is frequently assumed to be transferable to other systems such as elemental semiconductors \cite{C-W-Fabian-2010}, whereas proof for that or a rigorous calculation of the actual dependence is still lacking.  All this makes truly predictive theoretical methods highly desirable to quantify the achievable spin relaxation times for delocalized electrons in group-IV materials.

In this letter, we present such a method, which is able to calculate spin relaxation times fully from first-principles 
without adjustable parameters and without restrictions concerning the nature of the band gap. This method represents a new reliable and unbiased way to calculate spin relaxation times where eigenstates and phonon dispersions are calculated self-consistently and especially adds to the previously existing body of work the capability of calculating parameter-free the effect of arbitrary impurities or defects on spin-transport. 

We focus on the EY mechanism, which dominates the technologically relevant temperature range above 100 K \cite{Lepine-1970}. In the EY mechanism, spin relaxation is induced by momentum scattering off impurities or phonons. 
Within the Born approximation, the EY spin relaxation time can be related to the momentum relaxation time 
(which is proportional to the carrier mobility) \cite{Fabian-DS-1999}. The underlying theory to connect them exists on a phenomenological level for III-V semiconductors with direct gap \cite{Chazalviel-1975,Fishman-L-1977}, but not within a first principles framework that includes indirect band gap semiconductors \cite{Z-F-DS-2004}.
 
Since a methodology based on density-functional theory (DFT) to calculate electron mobilities (and thus momentum relaxation times) has been recently developed by one of us \cite{Restrepo-V-P-2009}, what is left to show here is establishing the relationship between the spin-flip and  momentum scattering matrix elements. Other than most previous work, we do not employ a semiempirical k$\cdot$p representation of the band structure to model the effect of spin-orbit coupling on the electronic wave functions \cite{C-A-B-Fishman-2001},  but rather use the spin-dependent DFT wave functions directly.

In the presence of spin orbit coupling, Bloch states are given by a mixture of spin-up and spin-down states \cite{Elliott-1954},
\begin{eqnarray}
\Psi_{{\bf k} n \Uparrow }({\bf r}) & & =\left[ a_{{\bf k} n }({\bf r})|\uparrow\rangle +b_{{\bf k} n}({\bf r})|\downarrow\rangle\right] e^{i{\bf k}\cdot{\bf r}}\nonumber\\
\Psi_{{\bf k} n \Downarrow }({\bf r}) & & =\left[ a_{-{\bf k} n }^*({\bf r})|\downarrow\rangle -b^*_{-{\bf k} n}({\bf r})|\uparrow\rangle\right] e^{i{\bf k}\cdot{\bf r}}
\label{eq1}
\end{eqnarray}
with lattice momentum {\bf k}, band index $n$, and effective spins or pseudospins $\Uparrow$ and $\Downarrow$. Using Eq.~(\ref{eq1}) for potentials that are slowly varying in space on the scale of the unit cell 
\cite{Chazalviel-1975,Fishman-L-1977}, the relationship between the spin-flip matrix elements and the momentum matrix elements becomes
\begin{eqnarray}
\langle \Psi_{{\bf k} n \Uparrow }| V_i| \Psi_{{\bf k}' n' \Downarrow}\rangle =\sum_{\bf G} \{ -a^*_{{\bf k},n}({\bf G})b^*_{{\bf -k'},n'}({\bf G}) +\nonumber\\
 b^*_{{\bf k},n}({\bf G})a^*_{{\bf -k'},n'}({\bf G}) \} \langle \Psi_{{\bf k} n \Uparrow }| V_i| \Psi_{{\bf k}' n' \Uparrow}\rangle ,
\label{eq2}
\end{eqnarray}
with $a_{{\bf k},n}({\bf G})$ and  $b_{{\bf k},n}({\bf G})$ being the Fourier transforms of $a_{{\bf k}, n }({\bf r})$ and $b_{{\bf k}, n }({\bf r})$ from Eq.~(\ref{eq1}).
 $V_i$ is a scattering operator which can refer here to either electron-phonon (i.e. lattice) or impurity scattering. Explicit expressions for these two types of scattering mechanisms will be given below. We assume there is no spin imbalance between up and down electrons. The momentum relaxation times can be calculated using DFT  and density functional perturbation theory (DFPT) \cite{Baroni-etal-2001}  as shown in \cite{Restrepo-V-P-2009}. The $a_{{\bf k}n}({\bf G})$ and $b_{{\bf k}n}({\bf G})$ coefficients are obtained from a DFT calculation 
that includes spin-orbit coupling. 

The average spin relaxation time $T_1$ is given by \cite{Yafet-1963,Ferry-2001}
\begin{equation}
\langle T_1\rangle =
\frac{\sum_n\int_0^{k_F}[T_1({\bf k},n)\: \frac{\partial f}{\partial \varepsilon}|_{\varepsilon=\varepsilon_F}\:d^3k ]}{\sum_n\int_0^{k_F}[\frac{\partial f}{\partial \varepsilon}|_{\varepsilon=\varepsilon_F}\: d^3k]}.
\label{eq3}
\end{equation}
where $\varepsilon=\varepsilon_{{\bf k},n}$ are the energy bands, $\varepsilon_F$ the Fermi energy, and $f(\varepsilon_{{\bf k},n})$ the Fermi function. For phonon scattering, $T_1({\bf k},n)$  is given by \cite{Yafet-1963}
\begin{eqnarray}
\frac{1}{T_1({\bf k},n)}=\frac{4\pi}{\hbar} & & \sum_{{\bf q}\lambda n'}|g^{{\bf q}\lambda}_{{\bf k+q}n'\Uparrow,{\bf k}n\Downarrow}|^2 \nonumber\\
\times \{ [f(\varepsilon_{{\bf k+q},n'}) + n_{{\bf q}\lambda}] & & \delta(\varepsilon_{{\bf k},n} - \varepsilon_{{\bf k+q},n'} + \hbar\omega_{{\bf q}\lambda} ) \nonumber\\
 + [1 + n_{{\bf q}\lambda} - f(\varepsilon_{{\bf k+q},n'}) ] & & \delta(\varepsilon_{{\bf k},n} - \varepsilon_{{\bf k+q},n'} - \hbar\omega_{{\bf q}\lambda} ) \},
\label{eq4}
\end{eqnarray}
with phonon energies $\omega_{{\bf q}\lambda}$ and Bose-Einstein occupation factors $n_{{\bf q}\lambda}$. The temperature dependence of $T_1$ is given by the latter two functions. $\bf q$ and $\lambda$ are the phonon wavevector and polarization. The spin-flip electron-phonon coupling function $g$ is given by \cite{Grimvall-1981}
\begin{eqnarray}
|g^{{\bf q}\lambda}_{{\bf k+q}n'\Uparrow,{\bf k}n\Downarrow}|& & =\sqrt{\hbar/(2M\omega_{{\bf q}\lambda})} 
\nonumber\\
\times\langle \Psi_{{\bf k+q} n' \Uparrow }| & & \frac{dV_{e-p}}{du_{{\bf q}\lambda}}\cdotp \varepsilon_{{\bf q}\lambda} | \Psi_{{\bf k} n \Downarrow}\rangle 
\label{eq5}
\end{eqnarray}
with atom mass $M$, phonon polarization vectors $\varepsilon_{{\bf q}\lambda}$ and atom displacements $u_{{\bf q}\lambda}$ from their equilibrium positions. The electron-phonon interaction potential $V_{e-p}$ is given by Eq. 19.1 of ref. \cite{Yafet-1963}
\begin{eqnarray}
V_{e-p}({\bf r},{\bf p},\sigma,{\bf R+u_{{\bf q}\lambda}})=V({\bf r},{\bf R+u_{{\bf q}\lambda}})
\nonumber\\
+\frac{\hbar}{4m^2c^2}\nabla_{\bf r}V({\bf r},{\bf R+u_{{\bf q}\lambda}})\times {\bf p}\cdotp {\bf \sigma}
\label{eq6}
\end{eqnarray}
where the first term is the screened one-electron potential $V$ that depends on the electronic position ${\bf r}$, equilibrium atomic position ${\bf R}$ and displacement $u_{{\bf q}\lambda}$. The second term is the spin-orbit potential ($V_{\rm SO}$). ${\bf p}$ is the electron momentum and ${\bf \sigma}$ are the spin Pauli matrices.
Force constants and phonon frequencies are computed from first principles using DFPT \cite{Baroni-etal-2001}. 

The inverse spin relaxation time for a density $n_d$ of defects or impurities is given by
\begin{eqnarray}
& & \frac{1}{T_1({\bf k},n)}=\frac{4\pi n_d}{\hbar}\sum_{n'}\frac{V}{(2\pi)^3} \nonumber\\ 
\times \int d^3k'| & & M_{nn'}({\bf k}\Uparrow, {\bf k'}\Downarrow)|^2 \delta(\varepsilon_{n'}({\bf k'}) - \varepsilon_n({\bf k}) ), 
\label{eq7}
\end{eqnarray}
where $V$ is the volume of the supercell. Within the Born approximation the spin-flip scattering matrix element is given by $M_{nn'}({\bf k}\Uparrow,{\bf k'}\Downarrow)=\langle n{\bf k}\Uparrow|\Delta V|n'{\bf k'}\Downarrow \rangle$,  where the self-consistent scattering potential $\Delta V$ is the difference between the potentials of a system with a defect or impurity and a reference "unperturbed" system. Electronic screening and computational issues due to the use of supercells have been addressed as in \cite{Restrepo-V-P-2009}. All necessary DFT and DFPT calculations were performed within the local density approximation in Quantum-ESPRESSO \cite{Baroni-etal-2010},  using norm-conserving pseudopotentials that explicitly include spin-orbit coupling \cite{Hemstreet-F-N-1993,Theurich-H-2001}. We have used 40x40x40 and 48x48x48 k-point meshes for silicon and diamond, respectively and 48x48x16 for graphite, which was the maximum possible with our computational setup. Convergence was checked by using lower meshes. We find that for silicon the convergence error is less than 1 $\%$, for diamond 5 $\%$, and for graphite 10 $\%$.

Since the spin-orbit coupling enters the equations directly through the wave functions and not through a combination of band gap and spin-orbit splitting parameters, it is a priori unclear to what extent the known local density approximation (LDA) band-gap problem is a problem for the accuracy of our parameter-free method (LDA predicts band gaps significantly smaller than experiment \cite{Windl-2004}). Our results show that, for the cases where we can compare to experiments, even having a smaller LDA band gap leads to good agreement with experimental data.
By applying pressure we are able to theoretically  modify the direct and indirect band gaps in silicon. The ab-initio pressure coefficients thus obtained ($dE_g/dP=-1.5$ and 0.3 meV/kbar for indirect and direct band gaps, respectively) are in very good agreement with experiments \cite{Turton-1999}.
Artificially changing the band gap by varying the lattice strain 
led to a slow dependence $(E_g^{0.67})$ of the spin flip amplitude as a function of absolute gap in silicon. This result gives the opposite trend and its dependence is much weaker than the predicted scaling of $E_g^{-2}$ given by perturbation theory applied to direct-bandgap semiconductors \cite{Z-F-DS-2004}, which in the literature has been assumed to be correct for Si as e.g. stated in Ref. \cite{C-W-Fabian-2010}. In contrast, we found in the case of diamond a dependence of $E_g^{-2.25}$. The opposite trend of band gap as a function of pressure in the case of diamond is related to the absence of $d$ states with the same quantum number as the low lying $s$ and $p$ orbitals in carbon \cite{Fahy-Cohen-1987}. In silicon the $d$ states mix with the $s$ and $p$ states thus lowering the energy of the conduction band near the $X$ point. With increased pressure, these mixed states go down in energy and thus give a negative pressure coefficient. This mixing does not happen in diamond. As a result, the pressure coefficient near the bottom of the conduction band is positive.

A calculation of $T_1$ for silicon due to phonon scattering using empirical pseudopotential and bond charge models has been reported recently by Cheng et al.~\cite{C-W-Fabian-2010}. Cheng's results agree very well with experiments. Additionally, two very recent papers \cite{Li-Dery-2011,Tang-Collins-Flatte-2012} used analytical models to describe the symmetry of the electron spin-phonon interactions in silicon in detail. In contrast to that, the use of fully first-principles DFT and DFPT here allows truly predictive parameter-free calculations and additionally enables calculation of  impurity scattering effects, for which no previous work exists. 

\begin{figure}
\includegraphics[width=1.0 \columnwidth, angle=-0, scale=0.9]{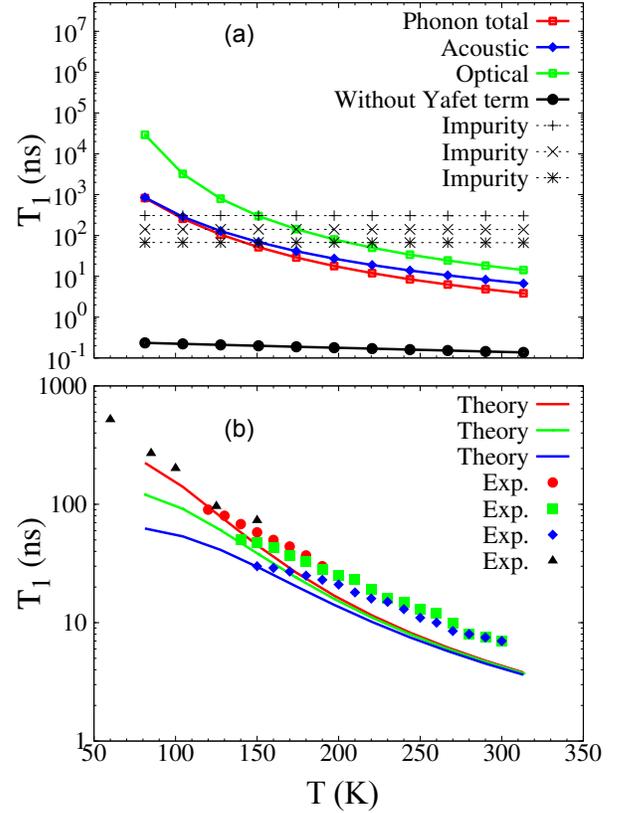} 
\caption{(Color online) (a) Phonon contribution to the electronic spin relaxation time $T_1$ in Si for all (red squares), acoustic (blue diamonds) and optical (green squares) phonons with Yafet term, and all phonons without Yafet term (black circles).
Dashed curves are contributions to $T_1$ from 
P$^+$-impurity scattering at carrier concentrations of $7.4\times 10^{14}$  ($+$), $7.8\times 10^{15}$ ($\times$) and $8\times 10^{16}$ cm$^{-3}$ ($*$). 
(b) Theoretical $T_1$ (lines) for Si compared with experimental values \cite{Lepine-1970} for carrier concentrations  of $7.4\times 10^{14}$ (circles), $7.8\times 10^{15}$ (squares), and $8\times 10^{16}$ cm$^{-3}$ (diamonds) and with \cite{Huang-M-A} (triangles).
}
\label{fig:Fig1} 
\end{figure}

$Silicon-$ Our results for the separate contributions to the spin relaxation time from acoustic and optical phonon scattering are shown in Fig. \ref{fig:Fig1}(a). Acoustic scattering is the most relevant phonon scattering mechanism throughout most of the temperature range. In contrast to previous suggestions \cite{Yafet-1963}, we however find that optical scattering starts to become competitive near room temperature. A $T^{-3.5}$ temperature dependence is found for phonon scattering. We find that the inclusion of the spin-orbit term (Yafet term) in the electron-phonon coupling function (second term in Eq.~(\ref{eq6})) is crucial for obtaining the right temperature dependence as was first pointed out by Yafet \cite{Yafet-1963}. Indeed, the cancellation of the zero and first order terms between the gradient of the electrostatic potential and the gradient of the spin-orbit potential leads to a higher $T_1$ and much stronger temperature dependence. Without the Yafet term (black curve in Fig. \ref{fig:Fig1}(a)), $T_1$ follows a $T^{-0.35}$ dependence which leads $T_1$ to be $\sim$3500 times smaller at 80 K and $\sim$30 times smaller at room temperature. The calculated temperature dependence of $T^{-3.5}$ lies between the high temperature limit of $T^{-2.5}$ and low temperature limit of $T^{-5}$ calculated by Yafet for acoustic phonons. Also in Fig. \ref{fig:Fig1}(a) we show the results for spin relaxation times for impurity scattering as a function of temperature for different carrier concentrations. Increasing the carrier concentration by adding more donors in the examined range makes the impurity scattering mechanism to become relevant at temperatures below 150 K, lowering the total spin relaxation time as expected. 

We have used Matthiessen's rule to add the lattice and impurity contributions to the total spin relaxation time. In Fig. \ref{fig:Fig1}(b) we compare our results with the ESR experiments of Lepine and Appelbaum's recent measurements. We find very good agreement in the region above 150 K, where the EY mechanism is considered to be dominant. At room temperature the calculated relaxation time of $T_1=4.3$ ns is well in the usable range~\cite{C-W-Fabian-2010}. 

To check if our method would also produce useful results in the degenerate-doping regime, we also calculated the room-temperature spin relaxation time for a donor concentration of $1.8\times 10^{19}$ cm$^{-3}$, for which Dash et al.  
recently suggested a lower bound for the electronic spin lifetime of initially 140 ps \cite{Dash-etal-2009}, then 285 ps \cite{Dash-2011}, nearly two orders of magnitude smaller than Lepine's results in the low-doped regime. This indicates the relevance of impurity scattering in lowering the spin lifetime in the highly-doped regime, which is important to keep depletion zones reasonably small. We find a value of 180 ps, in the range of the experimental values. From this comparison, it seems that the changes in band structure (impurity bands) and wave functions from degenerate doping \cite{Kodera-1966}  play a lesser role for the value of the spin relaxation time, since they are not included in our theory. We also find a weak dependence of the spin-mix amplitude as a function of band gap.  

\begin{figure}
\includegraphics[width=1.0 \columnwidth]{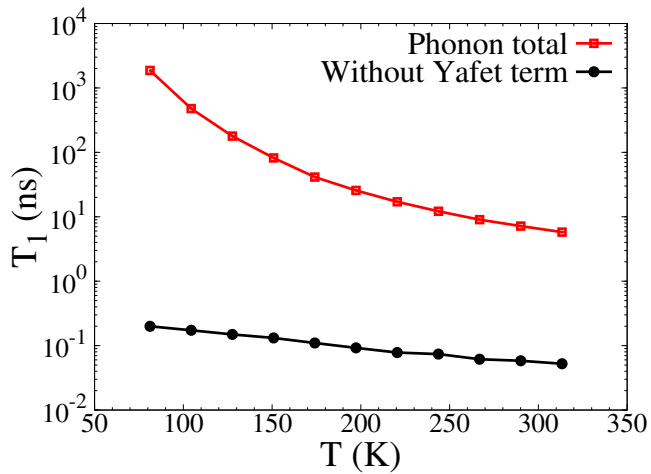} 
\caption{(Color online)
Spin relaxation due to phonon scattering in graphite. Red squares (black circles) were calculated with (without) inclusion of the Yafet term.}
\label{fig:Fig2} 
\end{figure}

$Graphite-$ The metallic character of graphite requires integration with special attention to the Fermi line that goes from the $K$ to the $H$ point of the Brillouin zone \cite{Boettger-1997} and inclusion of a minimum of two bands into the calculations. We obtain a spin relaxation time of $\sim$5.8 ns at 300 K (Fig. \ref{fig:Fig2}). This result is at the lower end of the available experimental data \cite{Wagoner-1960,Muller-1961} which report values between 1 and 300 ns. We find $T_1$ to be strongly anisotropic. By just considering scattering along the c axis, we get a much lower $T_1$ of 0.1 ns. We obtain a strong temperature dependence of $T^{-4.5}$, close to the theoretical low-temperature limit \cite{Yafet-1963}, as expected from the large in-plane Debye temperature of 2500 K. Without the Yafet term, $T_1$ would be $\sim$110 times smaller at room temperature and follow a linear temperature dependence.

\begin{figure}
\includegraphics[width=1.0 \columnwidth  ]{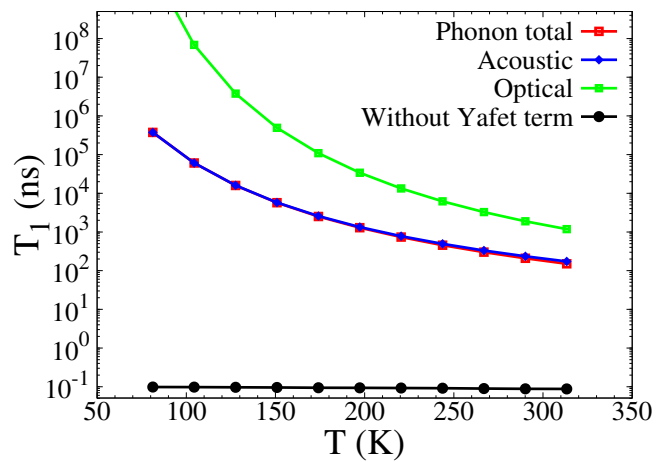} 
\caption{(Color online)
Calculated phonon contribution to the electronic spin relaxation time in diamond from all (red squares), acoustic (blue diamonds) and optical (green squares) phonons. The black circles show spin lifetimes calculated without the Yafet term. }
\label{fig:Fig3} 
\end{figure}

$Diamond-$ Figure \ref{fig:Fig3} shows the calculated $T_1$ for diamond considering only phonon scattering. As expected, diamond has a considerably larger $T_1$ than silicon with a value of 180 ns at room temperature. However, it is significantly smaller than the $10^2-10^4$~s speculated recently \cite{Balasub-2009}  using a formula \cite{Pake-1962} that in the case of silicon gives $T_1=1$ s at room temperature, eight orders of magnitude larger than experiment, as had already been cautioned in Ref.~\cite{Pake-1962}. This formula is based on the Waller theory of relaxation by modulation of dipole-dipole interactions by lattice vibrations \cite{Waller-1932}. This mechanism is very weak and typically results in spin relaxation times many orders of magnitude larger than measured experimentally \cite{Pake-1962}. On the experimental side, data are available for the dephasing times for the well-studied nitrogen-vacancy center in diamond ($\mu$s to ms) \cite{Balasub-2009,H-G-Awschalom-2006}, but an experimental study of the spin relaxation of conduction electrons is still lacking. In order to check if the magnitude of our results is reasonable, we have performed mobility calculations \cite{Restrepo-Windl} of conduction electrons at room temperature obtaining a value of 130 cm$^2$/Vs, which is within the range of recent experimental Hall data \cite{Pernot-etal-2006} (100-660 cm$^2$/Vs). 
Since $T_1$ for Si agrees well with experimental data and the electron mobility for Si has been accurately predicted previously \cite{Restrepo-V-P-2009}, the link between the matrix elements that deliver the mobility and the spin-relaxation time (Eq.~(\ref{eq2})) has been validated. Based on this chain of benchmarks, we are confident that the predictions for $T_1$ in diamond and graphite are sensible, and will also be reliable for other materials where spin relaxation is dominated by the same processes.
As temperature decreases, the $~ T^{-5}$ temperature dependence in diamond (expected as in graphite due to the high Debye temperature of 2200 K) leads to a much larger $T_1$ than in silicon ($\sim$110 times larger at 150 K). Accordingly, scattering from acoustic phonons is dominant throughout the whole temperature range up to room temperature. When we do not include the Yafet term in the calculation, $T_1$ becomes $\sim$2300 times smaller at 300 K and with almost no temperature dependence $(\sim T^{-0.07})$, illustrating again the relevance of including this term in the electron-phonon coupling function. Since the calculated band gap dependence of the spin-mix amplitude is $E_g^{-2.25}$ and the LDA band gap is 20 $\%$ smaller than the experimental gap, we estimate the true $T_1$ to vary within a factor of 1-1.6 from the reported ab-initio values.

In conclusion, a new parameter-free first-principles method to obtain spin relaxation times for phonon {\it and} impurity scattering has been presented, which is generally applicable to arbitrary systems. We have benchmarked this method for the Elliott-Yafet dominated temperature regime in silicon with very good agreement with experiment for phonon and impurity scattering and a limit for $T_1$ of 4.3 ns at room-temperature. For silicon, we find a weak dependence of the spin-mix amplitude as a function of band gap. We predict a stronger $T^{-5}$ temperature dependence for phonon scattering in diamond with $T_1 = 180$~ns at room temperature. This value is significantly smaller than the $10^2-10^4$~s recently estimated \cite{Balasub-2009} using a formula \cite{Pake-1962} that finds equally for Si a value that is eight orders of magnitude higher than experiment. Although no experimental results exist for $T_1$,  electron mobility measurements support our results. For graphite we find a $T^{-4.5}$ temperature dependence with a $T_1$ of 5.8 ns, at the lower end of the experimental range. 

We thank J. Sinova for valuable feedback.
This research was funded by the Center for Emergent Materials at The Ohio State University, an NSF MRSEC (Award Number DMR-0820414) and in part by NSF Award DMR-0925529. We also thank the Ohio Supercomputer Center for support under project \# PAS0072. 

\bibliographystyle{prsty}
\bibliography{spin_relax}

\noindent\\  

\end{document}